# Revealing the microstructure of sodium-montmorillonite aqueous suspensions

Mohammad Shoaib[a], Shaihroz Khan[a], Omar B Wani[a], Jitendra Mata[b], Anthony J. Krzysko[c], Ivan Kuzmenko[c], Markus Bleuel[d], Lindsey K. Fiddes[e], Eric W. Roth[f], Erin R Bobicki[a]

[a]Department of Chemical Engineering and Applied Chemistry, University of Toronto, Canada Ontario, M5S 3E5, Canada

[b]Australian Centre for Neutron Scattering (ACNS), Australian Nuclear Science and Technology Organisation (ANSTO), Lucas Heights, Sydney, NSW 2232, Australia

[c]Argonne National Laboratory, 9700 S. Cass Avenue, Bldg. 433A, Argonne, Illinois 60439, United States

[d]NIST Center for Neutron Research, National Institute of Standards and Technology, Gaithersburg, Maryland 20899, United States

[d]Department of Materials Science and Engineering University of Maryland College Park, MD 20742-2115

[e]Temerty Faculty of Medicine, University of Toronto, Toronto, Ontario

[f]Northwestern University Atomic and Nanoscale Characterization Experimental Center, Northwestern University, 2220 Campus Drive, Evanston, Illinois 60208, United States

**Abstract**

Aqueous suspensions of geometrically anisometric (2D) sodium-montmorillonite (Na-Mt) particles display a sol–gel transition at very low solids concentrations. The underlying microstructure of the gel has remained a point of contention since the time of Irving Langmuir. An in-situ investigation encompassing length scales much larger than the individual particles is required to provide support for one of the two models proposed in the literature: 1) a percolated network governed by electrostatic attraction between platelets; and 2) a jammed suspension stabilized by repulsive electrostatic forces between particles. We settle this debate by comprehensively probing the microstructure of Na-Mt suspensions using ultra-small angle neutron/X-ray scattering and found that it is ordered and contains entities that are at least an order of magnitude larger than the individual particles. Complementary cryo-electron microscopy showed both the presence of domains having strong particle-particle ordering and regions of particle-particle aggregation. These data indicate 1) the presence of nematic domains, which refutes a purely attractive nature, and 2) assembly of particles, which refutes a purely repulsive nature. Na-Mt gels appear to have a hybrid microstructure with both attractive and repulsive domains.

**Introduction**

Montmorillonite is a naturally occurring phyllosilicate mineral (i.e., clay) that plays an important role in natural processes like debris flows and industrial systems including drilling, paper making, nuclear waste disposal, health and beauty product manufacturing, and mineral processing[1-5]. The colloidal behaviour of sodium-montmorillonite (Na-Mt) has been the subject of numerous studies dating back to the seminal work of Irving Langmuir and Herbert Freundlich[6, 7]. Na-Mt consists of a dioctahedral aluminium hydroxide sheet sandwiched between two tetrahedral silica sheets. Isomorphic substitutions by less charged cations produce a net negative layer charge, but it is the valency and hydration properties of interlayer exchangeable cations that control both swelling and colloidal behaviour[8-11].

Unlike suspensions of swelling clays such as nontronite (NAu-1), nontronite (Nau-2), and beidellite (SBId-1) that display a clear entropy-driven, first-order isotropic to nematic transition[12-16], Na-Mt suspensions display a sol-gel transition at very low volume fractions[17]. The existence of gelation prevents suspensions from reaching a true thermodynamic equilibrium[18, 19]. The microstructure of the gel has remained a point of contention in the literature. Two models have been proposed based on either 1) the formation of a tridimensional network governed by electrostatic attraction between platelets[20, 21] or 2) an oriented network stabilized by repulsive forces caused by interacting double layers[22, 23, 24-27]. Whereas the sol state exhibits a shear-induced optical birefringence that disappears once shearing is stopped, the gel state remains permanently birefringent[28]. As a result, the development of yield stress beyond the gel point is attributed to a repulsive jamming transition rather than a percolation threshold.

The first model above is supported by the fact that, even in the sol state, the shear-induced ordering transition occurs at very low Peclet numbers based on the size of individual particles. At these low Peclet numbers, the particles should not be oriented and should be randomly organized. The time required for shear-induced ordering to disappear when the shear stress is removed is longer that the rotational time of a single particle[29]. Moreover, the number density of particles required for the sol-gel transition evolves as the inverse of the average particle diameter, suggesting that the entities that remain at the origin of the sol-gel transition are clusters or stacks, not individual particles[30].

After several decades of rigorous research, the precise microstructure of Na-Mt gels remains elusive. Although DLVO (Derjaguin, Landau, Vervey, and Overbeek) theory has played a major role in our understanding of colloidal suspensions, its applicability to swelling clay suspensions—similar to the case of cement where the DLVO theory does not predict any cohesion[31, 32]—is debated due to possible contributions of long-range attraction between like-charged particles [33-36]. Further, the issue is not fully resolved from a theoretical point of view[37, 38]. Furthermore, Batista et al.[39] highlighted the nonadditivity of electrostatic, van der Waals and other interactions in suspensions of nanoparticles especially < 100 nm resulting in breakdown of DLVO theory to understand the interaction in nano-suspensions. The unfractionated Na-Mt suspensions are a mix of particles with a size range from ~ 20-600 nm, therefore, due to the nonadditivity of interactions, the application of DLVO to understand this system is rather challenging. Previous studies have either been limited by the minimum scattering angle, which can only cover a length scale up to 1 µm, or they have relied on rheology to discern the microstructure. A more robust method would probe the microstructure at length scales much larger than individual particles, which requires the ability to measure very small angles[40]. Here, we probe the microstructure of Na-Mt suspensions below and above the gel point using ultra-small angle X-ray scattering (USAXS) and ultra-small angle neutron scattering (USANS) along with high-pressure cryogenic scanning electron microscopy (cryo-SEM), cryogenic transmission electron microscopy (cryo-TEM), and dilution studies on ultra-low ionic strength ($10^{-5}$ M) Swy-3 Na-Mt suspensions.

**Results and discussion**

The osmotic pressure of Na-Mt suspensions increased gradually as a function of concentration in the sol state (**Figure 1a**). Suspensions remained in the sol state until approximately $10^3$ Pa osmotic pressure (3 wt.% concentration). In the gel state, the osmotic pressure continued to gradually increase until 4.2 wt.%, above which it increased strongly, and it became extremely difficult to further concentrate the suspensions. Osmotic pressure represents the force resisting the extraction of liquid. It originates from an increase in energy (interparticle forces) and/or a loss of entropy (configurations)[41, 42]. The dramatic increase in the pressure required to reach concentrations above 4.2 wt.% signifies enhanced interparticle forces in the gel state. These results agree with other studies on Na-Mt systems[24, 28].

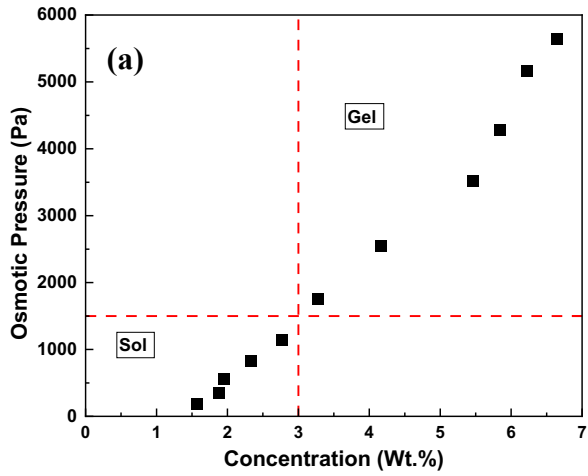
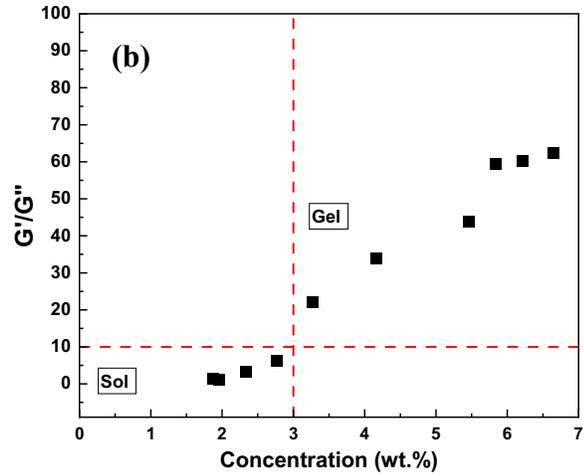
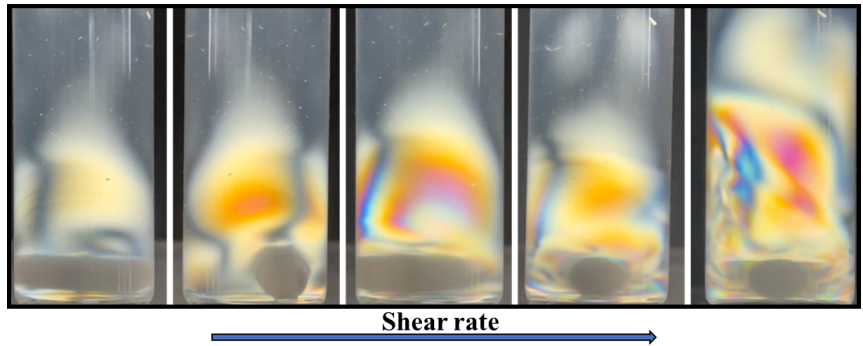
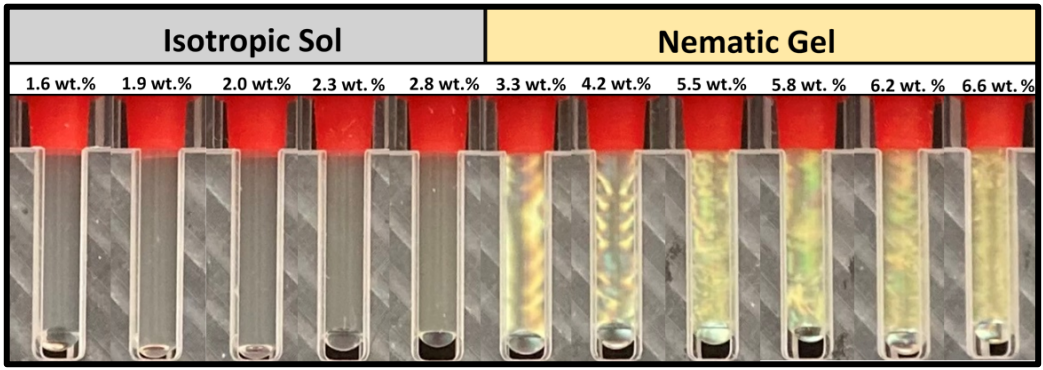
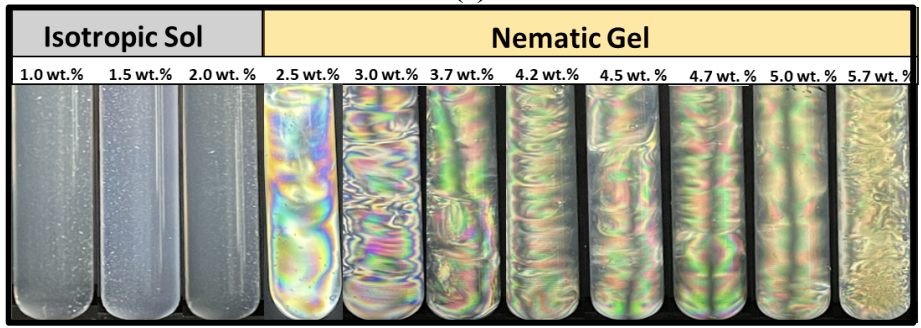

**Figure 1: (a) Osmotic pressure vs Swy-3 Na-Mt suspension concentration, (b) ratio of storage modulus (G') to loss modulus (G'') for Na-Mt suspensions in the sol and gel states, and (c) visualizations of Na-Mt suspensions in the sol and gel states between crossed polarizers. (d) shear induced birefringence in sol state of Na-Mt suspensions, the birefringence grows as the shear rate is increased signifying enhanced ordering at higher shear rates(e) visualizations of Na-Mt suspensions in Deuterium oxide ($D_2O$), in the sol and gel states between crossed polarizers. Note that the birefringent pattern evolves as a function of concentration in the gel state. The gel point is between Na-Mt concentration 2.8-3.3 wt.% in $H_2O$ and between 2.0-2.5 wt.% in $D_2O$.**

The ratio of the linear viscoelastic storage modulus (G') to loss modulus (G'') was below 10 in the sol state and above 10 beyond the gel point (**Figure 1b**). Similar to osmotic pressure, the ratio increased gradually as a function of concentration in the sol state, but unlike osmotic pressure, the ratio increased more gradually with increasing Na-Mt concentration in the gel state. In the gel state, G' evolves exponentially similar to the osmotic pressure indicating a common underlying mechanism responsible for osmotic pressure and G' in Na-Mt suspensions. The G' acquires values of the order of $10^3$ Pa at higher concentrations signifying a very high strength of the gelled microstructure (**Figure S2a**). Similar results are also obtained for Na-Mt in deuterium oxide ($D_2O$) except that the gel point is achieved at lower Na-Mt weight concentration in the presence of $D_2O$ due to the density difference between $D_2O$ and water which meant higher volume fraction of Na-Mt at the same weight concentration (**Figure 1d, Figure S2 b, c**).

When viewed between crossed polarizers, Na-Mt suspensions in the sol state remained optically isotropic and only briefly exhibited birefringence immediately after being sheared (**Figure 1c**), whereas in the gel state, they exhibited a permanent birefringence or nematic character (**Figure 1d**). The time required for the shear-induced birefringence to disappear increased as a function of concentration in the sol state until the system "locked" the particles or entity responsible for the nematic character at the gel point. Therefore, the sol and gel states both exhibited ordering. For the sol state, an external shear was required for the ordering to be evident, and for the gel state, a mechanism was needed through which the shear-induced effect could be retained. Therefore, the relaxation time—the time required for the microstructure to transition from order to disorder—was very short in sol state and very long in the gel state[29]. This shear-induced and permanent birefringence has also been observed in aqueous suspensions of other geometrically anisotropic particles (e.g., laponite and cellulosic nanocrystal)[29, 43]. The main question is how the driving force differs between the shear-induced ordering in the sol state and the permanent ordering in the gel state.

**Ultra Small Angle X-Ray Scattering (USAXS):** The slope of the intensity versus scattering wave vector (Q) curve was approximately –2 (**Figure 2 a, b**), signifying the nearly bidimensional nature of the disks[14, 44]. The USAXS data at the lowest Q values in sol state can be seen to be hardly flattening up at the lowest three concentrations and takes an upturn above 2.3 wt.% concentration. From this concentration onward, the slope of the curve at the lowest Q values increases significantly which suggests the entities grow as a function of concentration[45-48]. Therefore, the USAXS results provided the proof that entities much larger than ∼6 μm are present in Na-Mt suspensions. Further, a "bump" in the vicinity of $10^{-2}$ Å$^{-1}$, which is more visible when $Q^2I$ is potted against Q (**Figure 2 c, d**), corresponds to the short-range positional order of the platelets or characteristic repeat distance in the underlying microstructure. This has been provided as evidence of a purely repulsive microstructure in Na-Mt suspensions[24, 49]. However, our results clearly show that there is much to the microstructure beyond the length scale where this ordering is observed. Therefore, the bump may correspond to the short-range positional order of one type of platelets that are smaller. A second bump in the curve at $4\times10^{-4}$ Å$^{-1}$ (~1.5 μm) may be the distance between aggregated entities composed of several larger particles. The microstructure can then be viewed as hybrid, containing segregated zones of ordered particles responsible for birefringence and aggregated particles responsible for gelation. Therefore, the microstructure of Na-Mt suspensions is complex rather than simply repulsive or attractive.

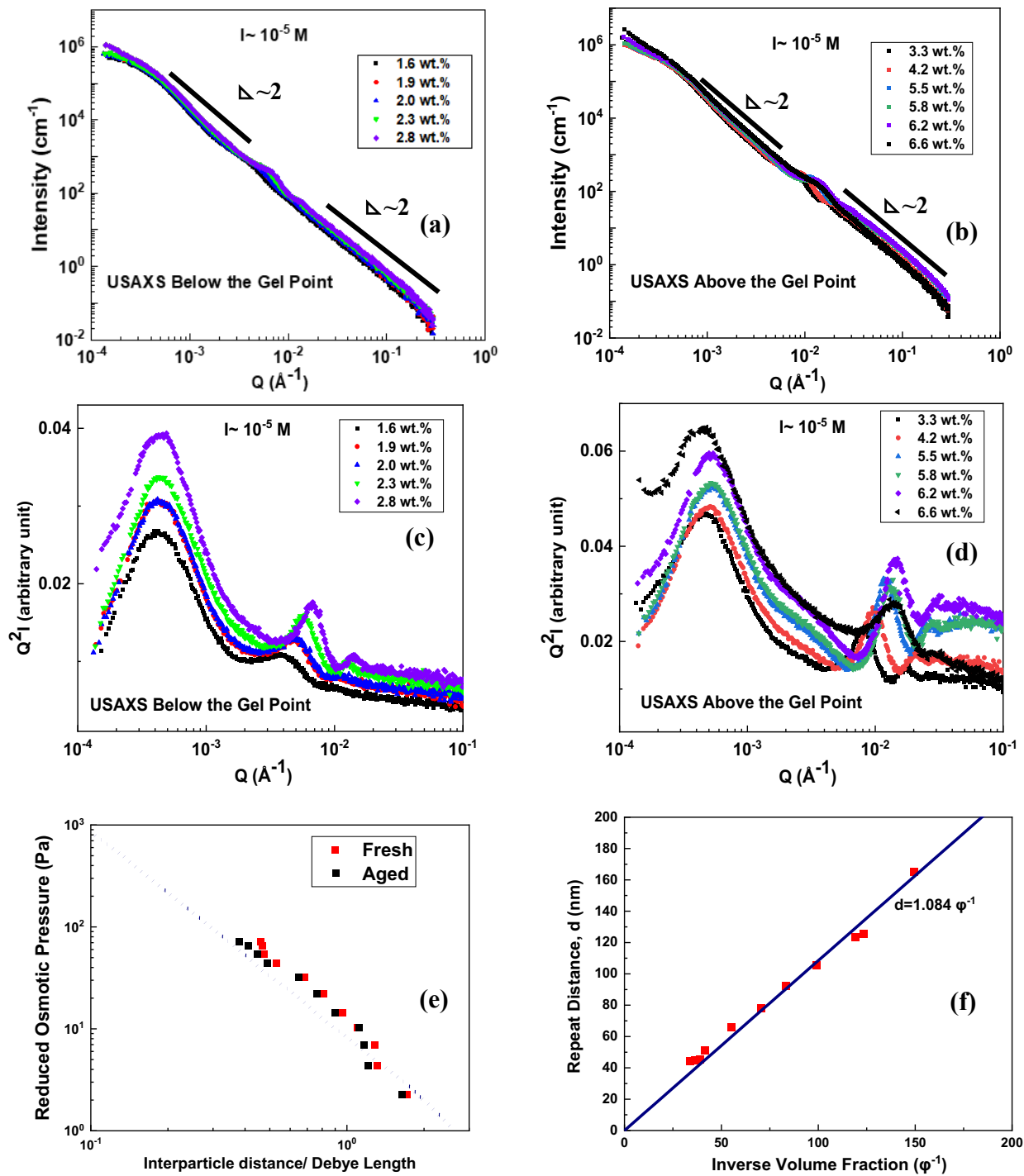

**Figure 2:** Ultra-small angle X-ray scattering analysis of low ionic strength ($10^{-5}$ M) sodium-montmorillonite suspensions (a,c) in the sol state and (b,d) in the gel state; (e) reduced osmotic pressure vs interparticle distance scaled to Debye length; dashed line represents asymptotic solution of eq A6 in [30] with σ = −0.11 C·m$^{-2}$ and H ≫ 0.3 nm for a

**symmetrical 1:1 electrolyte solution; and (f) evolution of repeat distance between particles vs inverse volume fraction.**

The osmotic pressure decayed exponentially as a function of average interparticle distance obtained from the length scale at maximum oscillation of the scattering intensity ($2\pi/Q$). The reduced osmotic pressure ($P_{osm}/4 \times C_{salt}RT$) plotted against the interparticle distance normalized to Debye length fell close to the analytical solution of Poisson-Boltzmann equation solved for a symmetrical 1:1 electrolyte for two parallel infinite plates at separation distances >>0.64 nm [24, 30] (**Figure 2e**). These results are similar to those reported for Swy-2 montmorillonite based on small angle X-ray scattering[24] and signify the role of repulsive electrostatic interactions in the microstructure. It is worth noting that a permanent birefringence pattern develops when the interparticle separation is equivalent to the Debye length ($d/k^{-1} \sim 1$), further confirming the role of electrical double layer interactions in the microstructure. Swelling law relating the interparticle distance to volume fraction approximately follows a unidimensional swelling for which the average thickness of the individual particles is obtained by $D = t\varphi^{-1}$, where t is the layer thickness[14, 24, 50] (**Figure 2f**). The slope of the line suggests the platelet thickness of particles responsible for the diffuse peak is approximately 1 nm, which is close to the thickness of a single clay sheet[24].

Microstructure evolution monitored over three months of aging with USAXS did not show any dramatic variation, apart from minute changes signifying very slow structural changes. These include enhanced aggregation as reflected by the extreme upturn in the scattering intensity at the lowest scattering angles at high Na-Mt concentrations in the gel state (**Figure 2d**) and a smaller interparticle distance in the aged than fresh system (**Figure 2e**). The aging results thus indicate that the Na-Mt suspensions were out of equilibrium and additional study focusing on this aspect using advanced techniques such as X-ray photon correlation spectroscopy is warranted.

The SAXS profiles for sol and gel state were similar with intensity decaying at a slope of ~ – 2 till Q values of around $\text{Å}^{o-1}$ (**Figure S4**). The SAXS region also didn't have peak in the vicinity of 0.1 Å$^{-1}$ which typically represents the presence of tactoids in Na-Mt suspensions[51]. The aging of samples didn't influence the SAXS profile at all (**Figure S6**). Similarly, WAXS region scattering for both the fresh and aged samples were also similar suggesting the crystal structure of the sample remained intact (**Figure S4 and S6**).

**Ultra Small Angle Neutron Scattering (USANS):** After confirming the presence of objects much larger than the scale covered by USAXS, we probed the length scale of the entities up to 17 μm using USANS. Both the sol (**Figure 3a**) and gel states (**Figure 3b**) exhibited scattering at the lowest scattering angles, indicating the presence of entities much larger than 17 μm[47]. The slope of the intensity vs q curve increased with concentration in both states, showing that aggregation of Na-Mt suspensions increased in extent with increasing concentration. These results further clarify the microstructure of Na-Mt suspensions and contradict a purely repulsive microstructure.

The effect of ionic strength on the microstructure was also studied with USANS. For the sol state, the introduction of salt resulted in a counterintuitive reduction in intensity but scattering at the lowest Q was similar regardless of the ionic strength, signifying aggregation. For the gel state, increasing ionic strength increased the intensity, signifying intensified aggregation in the presence of salt. The USAXS and USANS scattering data in the gel state covering a length scale of 1 Å to 17 μm overlapped, albeit with a slight vertical shift to account for the intensity difference between the two methods (**Figure 3e**). The bending in scattering curve around a Q of $4\times10^{-4}$ Å$^{-1}$ was consistent for both USAXS and USANS.

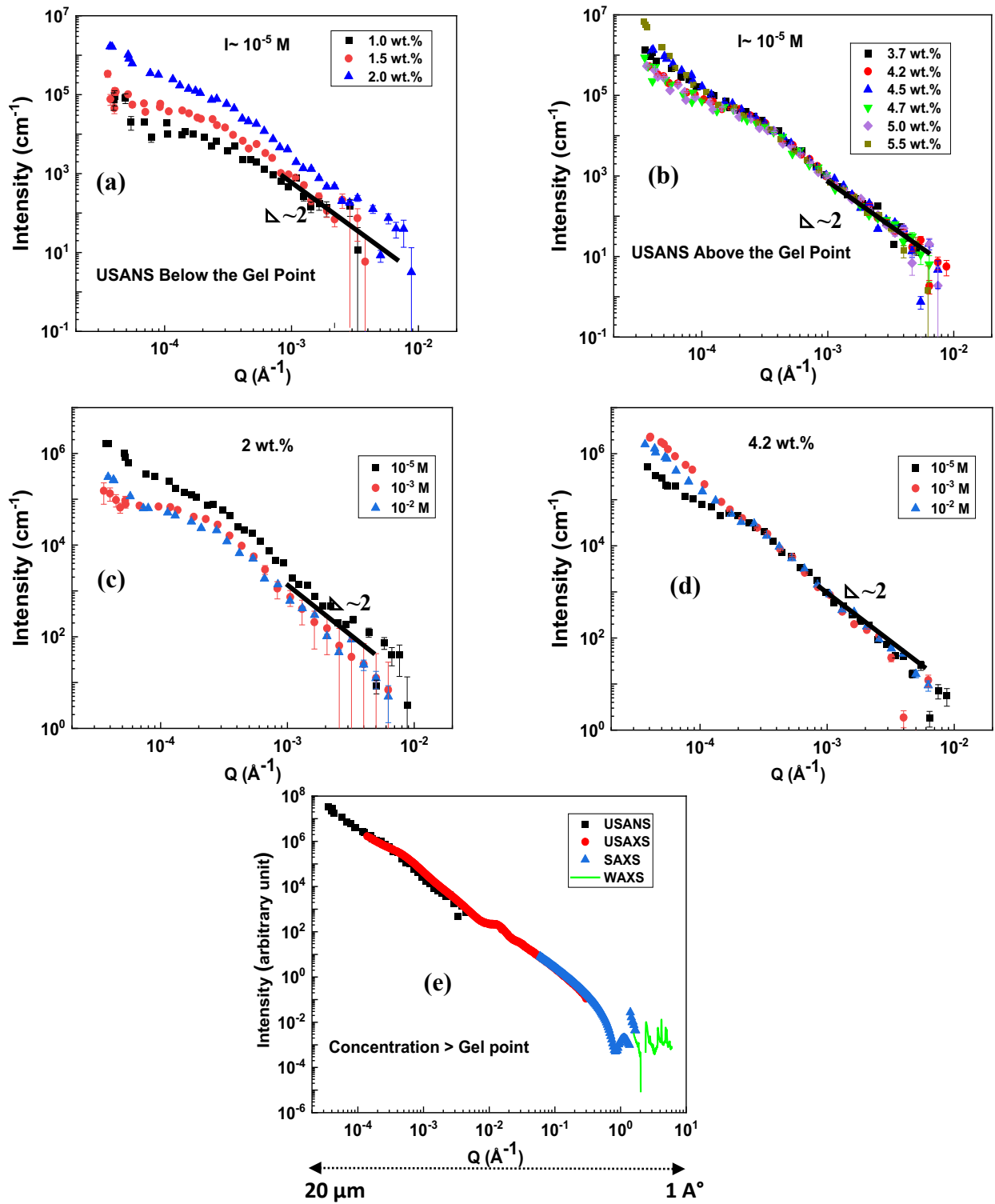

**Figure 3:** Ultra-small angle neutron scattering (USANS) of low ionic strength ($10^{-5}$ M) Swy-3 sodium-montmorillonite (Na-Mt) suspensions (a) in the sol state and (b) in the gel state; representative USANS data for Na-Mt suspensions (c) in the sol state and (d) and in the gel

**state at three ionic strengths; and (e) combined scattering data from wide (WAXS), small (SAXS), and ultra-small angle scattering (USAXS) and USANS in the gel state.**

**High-pressure cryo-SEM and Cryo-TEM:** The sol state exhibited few contacts between particles without any ordering (**Figure 4a, c**), whereas the gel state had regions of particle aggregation along with domains where particles were ordered in a face-face manner (**Figure 4b, d**). Unlike the plunge frozen technique during Cryo-SEM sample preparation, which produces a honeycomb structure[52, 53], the high-pressure technique used here produced no honeycomb pattern. These complementary results confirm the presence of aggregation in Na-Mt suspensions in both sol and gel states. It is worth noting that the interparticle distance measured on cryo-EM images is significantly greater than the interparticle distance (~ 15 Å) reported for tactoids[54]. Therefore, the nematic character of the gel is attributed to face-face ordering of the Na-Mt particles.

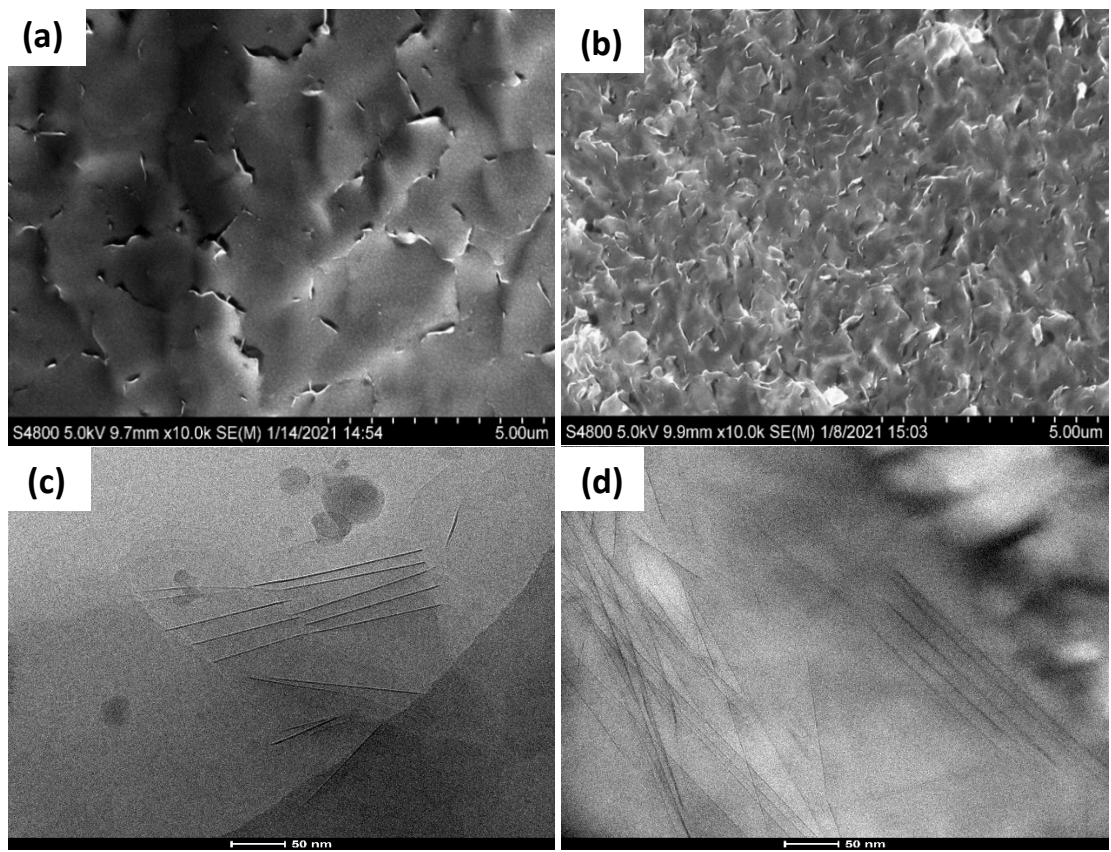

**Figure 4: Cryo-SEM (a,b) and -TEM (c, d) images of Swy-3 sodium-montmorillonite in (a, b) sol (1.9 wt.%) and (b, d) gel states (6.6 wt.%)**

**Dilution behaviour:** The dilution method used in previous research to distinguish between a repulsive and an attractive system was used here[55, 56]. The gel state showed a tendency to melt at all concentrations, with the time required to melt increasing with concentration (**Figure 5**). The time required for the gel to melt completely was significantly greater for Na-Mt than for laponite, a synthetic swelling clay[45]. Although aggregation in the system was confirmed, the system still melts when in contact with water. This may be linked to the transport characteristics of water inside the gel, which may be related to high-permeability streaks in the gel arising coming from strongly ordered domains. As a result, water penetrates easily into the structure at lower concentrations since the particles are not tightly packed. At higher concentrations, particles are closer together, and the structure has lower permeability and porosity.

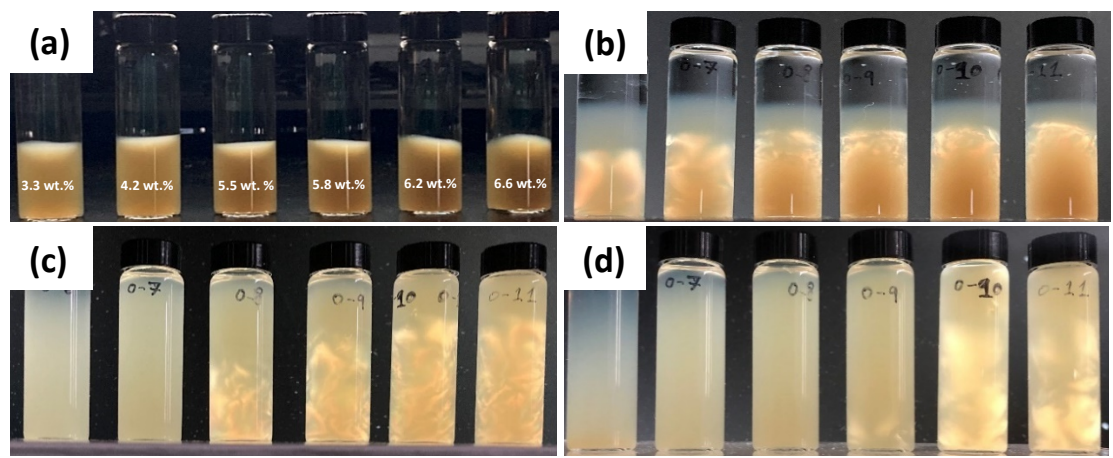

**Figure 5:** Dilution behaviour of Swy-3 sodium-montmorillonite suspensions at different concentrations in the gel state after (a) 0, (b) 2, (c) 47, and (d) 150 days.

## Discussion

We probed the microstructure of Na-Mt aqueous suspensions covering a length scale of more than five orders from interatomic distances using WAXS and to more than an order larger than the particle size using USAXS and USANS. The scattering profile in the WAXS region remained unaltered as a function of concentration; thus, the microstructure remained unaffected at these length scales. For the SAXS region, the intensity decayed at a rate of ~2, regardless of Na-Mt concentration. Previous work on Na-Mt suspensions covered a length scale up to only a maximum of 100 nm, which is within the particle size range of Na-Mt particles[51, 57-60]. Therefore, the presence of Bragg's peak in the SAXS region is evidence for a purely repulsive microstructure. Scattering at larger length scales using USAXS showed that the microstructure contains entities larger than ~6 μm, and the aggregation at this length scale only intensified as a function of concentration. USANS further showed that the aggregated entities are larger than 17 μm. Therefore, the presence of scattering objects at a length scale of 17 μm and Bragg's peaks in the SAXS region signify the presence of two domains: one has particle-particle aggregation, and the other has particle-particle ordering.

This was further confirmed by cryo-SEM and cryo-TEM images where particle-particle aggregation and particle-particle ordering were clearly visible. The presence of domains was also reflected in dilution behaviour, whereby the gelled structure melted upon contact with water. The time required to melt the gel increased as a function of concentration due to the enhanced gel strength and aggregation as a function of concentration. Therefore, the Bragg's peak in the SAXS region only reveals particle-particle ordering from unaggregated domains. Bihannic et al.[17] detected the existence of large-scale structures much larger than individual clay platelets formed by alternating clay-rich and clay-poor domains using synchrotron-based X-ray fluorescence microscopy. However, these authors used osmotically prepared suspensions, and the structure within the domains was not detectable due to the insufficient resolution of the technique.

The fundamental mechanism behind the formation of particle-particle ordering can be understood from Onsager's liquid crystal theory[61]. However, the presence of aggregated domains is perplexing and raises several fundamental questions regarding the mechanisms and properties of gels in systems of charged colloidal platelets. There may be one or a combination of mechanisms responsible for aggregation. Firstly, the Na-Mt particle size distribution is quite broad, and the surface charge characteristics of these particles may vary as a function of particle size range. The larger particle size range may have low-charge regions that can aggregate in a patch-wise manner through alignment of low- or opposite-charge domains (positive facing negative) especially at higher particle concentrations[62, 63]. Smaller particles may drive aggregation of larger particles due to depletion; this effect may be enhanced if the larger particle size range also has low-charge regions. If this is the case, then suspensions of fractionated Na-Mt particles will have a strong particle-size-dependent microstructure.

Michot et al.[28] reported the effect of particle size on the phase diagram of fractionated Na-Mt suspensions obtained by successive centrifugation. The authors concluded that the effect of anisotropy on Na-Mt suspensions was opposite to that of other systems of charged colloidal platelets, where smaller particles exhibited a sol-gel transition at lower solid concentrations than larger particles. The inverse relationship of sol-gel transition was speculated to be linked to particle-particle association, giving rise to disconnected clusters whose amount increases linearly with decreasing particle size. Such disconnected clusters were speculated to give rise to a heterogeneous suspension, which we have confirmed using scattering and cryo-EM experiments. Ion-ion association effects on the appearance of attraction between equally charged particles in the presence of multivalent ions (e.g., cement cohesion) should also be investigated and may drive the particle-particle aggregation observed here[31, 64-66].

**Conclusions**

The microstructure of Na-Mt suspensions was probed using USAXS, USANS, and complementary Cryo-electron microscopy and dilution studies. For the first time, the microstructure of these suspensions was studied up to a length scale of 17 μm, which is more than an order of magnitude larger than the largest particle size. Neither of the two models proposed to date for Na-Mt microstructure—a purely jammed suspension stabilized by repulsive electrostatic forces between particles and a percolated network governed by electrostatic attraction between platelets—accurately capture the real microstructure. The scattering results reveal the microstructure contains entities that are at least one order of magnitude larger than the individual particles. They also indicate the presence of ordering. They confirm the presence of domains having strong particle-particle ordering assembly of particles, refuting a purely repulsive nature for these gels. The presence of nematic domains refutes a purely attractive nature. The microstructure can be viewed as a hybrid that has both attractive and repulsive domains.

**Acknowledgements**

The authors wish to acknowledge Professor Daniel Bonn for his suggestion to use Neutron and X-ray scattering to probe the microstructure and Erwan Paineau for answering several questions related to the sample preparation.

**Materials and Methods**

**Sample Preparation:** Swy-3 montmorillonite clay was obtained from the Source Clays Minerals Repository of the Clay Mineral Society (Purdue University). The measured BET surface area was 34.02 m$^2$/g. The clay sample was purified by settling a 45 g/L suspension of clay in 1 M NaCl in an Imhoff cone for 72 h with three 1 M NaCl exchanges. The bottom fraction of the suspension was discarded, and the top fraction was dialyzed against deionized (DI) water at a water:suspension ratio of 50. The DI water was exchanged several times until the conductivity of the reservoir fell below 5 μS·m$^{-1}$, and a silver nitrate test confirmed the absence of chloride ions. This procedure resulted in a fully Na$^+$-exchanged montmorillonite at ultra-low ionic strength. The suspension obtained from the dialysis tube was centrifuged at 6,300 g for 90 min to remove coarse particles. The pellet was discarded, and the final stock suspension was collected in a beaker. The weight concentration of stock suspension was obtained by oven-drying a small sample at 120°C for 48 h. The final stock suspension was dialyzed in membranes with a molecular weight cut-off of 12,000–14,000 Da.

Osmotic stress experiments were performed with polyethylene glycol (molecular weight 35,000 Da) in DI water at different concentrations to obtain clay suspensions in sol and gel states. Osmotic stress experiments were equilibrated for 30 days, during which time the polymer solution was renewed three times. Final clay suspension concentrations were determined by weight loss upon oven-drying a small fraction of equilibrated suspensions at 120°C for 48 h. Osmotic pressure values for 35,000 Da polyethylene glycol can be found elsewhere[41].

USANS samples were prepared in D$_2$O obtained from Sigma Aldrich (Item # 7789-20-0). The dried sample obtained after centrifugation step was dispersed in D$_2$O at 0.50 wt.% and equilibrated for a month after which the sample was filled in membranes of molecular weight cut-off of 12,000–14,000 Da. A concentrated stock suspensions was obtained by applying osmotic stress using 35,000 Da polyethylene glycol in D$_2$O. The osmotic pressure was increased gradually from 1,000 to 8,000 Pa by increasing the concentration of polyethylene glycol. The concentrated stock suspension obtained was diluted using D$_2$O to obtain several concentrations of Na-Mt in D$_2$O.

**Particle Size Analysis:** The maximum particle size of the montmorillonite stock suspension measured with a Nanoparticle Tracking Analysis instrument (NanoSight NSss300, Malvern) by tracking more than 1000 particles (**Figure S1**) was lesser than 600 nm. Atomic force microscopy was also carried out to determine the size and thickness of purified clay particles. The surfaces of freshly cleaved, high-grade (V-1), 12 mm mica discs (Ted Pella Inc., Redding, CA) were treated for 5 min with 10 μL of 3-aminopropyl-trietoxy silane (1 μM in DI water), rinsed with 2 mL DI water, and blow dried. A 10 μL drop of 0.012 mg/mL montmorillonite suspension was incubated on the mica discs for 5 min in a wet chamber to avoid desiccation. Discs were dried in a laboratory fumehood at room temperature and scanned immediately on a BioScope™ II atomic force microscope (Bruker Corporation, Billerica, MA). High-resolution images of montmorillonite particles were obtained using RTESP cantilevers (fo=237–289 kHz, k=20–80 N/m, Bruker Corporation, Billerica, MA) (**Figure M1**). The particle nanotopography was determined using the tapping mode in air at a 0.7 Hz scan rate. Particle analysis, size distribution, and three-dimensional images were obtained with NanoScope Analysis© software (Ver. 1.50, Bruker Corporation, Billerica, MA). The particle size range obtained from the two methods were in good agreement with each other.

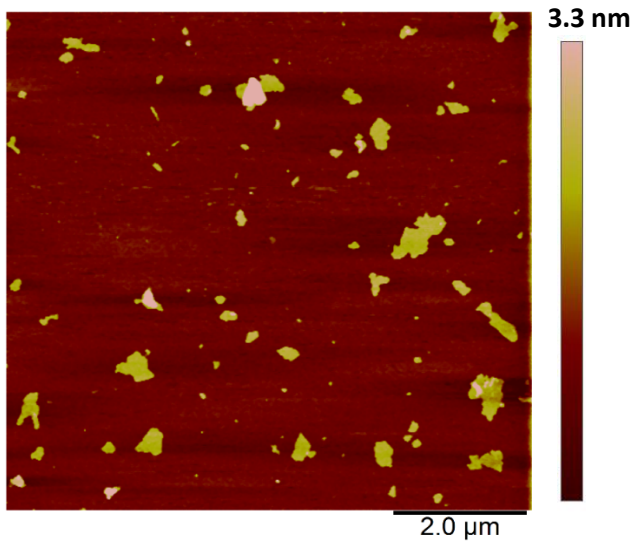

**Figure M1: Representative atomic force microscopy image of the Na-Mt sample**

**Mean diameter: 218 ± 127 nm (range 20–632 nm)**

**Mean thickness: 2 ± 0.824 nm (range 0.55–6.7 nm)**

**Rheology:** A stress-controlled Discovery HR-2 (TA Instruments) rheometer was used to measure the rheology of Na-Mt suspensions at 25°C. A concentric cylinder geometry (bob and cylinder wall sandblasted to limit wall slip) was used, with bob and cup diameters of 28 and 30.4 mm, respectively, resulting in a shearing gap of 1.2 mm. A thin layer of silicone oil was applied to the top surface of suspensions to prevent evaporation during tests. The overall procedure was as follows: 1) pre-shear at +1500 s$^{-1}$ and –1500 s$^{-1}$ for 2 min each to minimize the effect of sample loading history; and 2) wait 60 min and then perform the strain amplitude sweep at 1 Hz to obtain the linear viscoelastic region storage modulus (G') (**Figure S2**).

**Ultra-small angle x-ray scattering (USAXS):** Measurements were carried out at the 9-ID-C beamline of the Advanced Photon Source at Argonne National Laboratory[67]. Samples were placed in 1-mm capillary tubes, which were then sealed. The wavelength (λ) of the incident X-ray was 0.5904 Å (21 keV). The beam size was ≈ 0.8 (width) × 0.2 (height) mm$^2$ for SAXS and WAXS and 0.8 (width) × 0.8 (height) mm$^2$ for USAXS. Each USAXS measurement was acquired for 60 s with a scattering vector, Q (Q=4πsinθλ$^{-1}$)[68] ranged from $10^{-4}$ to $6 \times 10^{-1}$ Å$^{-1}$ with a resolution ≈ $8 \times 10^{-5}$ Å$^{-1}$. The SAXS and WAXS data were acquired for 30 s each over Q ranges of $3 \times 10^{-2}$–1.3 and 1–7 Å$^{-1}$, respectively. The background solution ($10^{-5}$ M NaCl in DI water) was subtracted from the main intensities. Data were reduced using the INDRA and NIKA[69] software packages and analyzed using the IRENA software package[70]. Data were also desmeared from slit smeared collimation of the Bonse-Hart USAXS system.

**Ultra-small angle neutron scattering (USANS):** The USANS measurements were performed on the KOOKABURRA[71] beamline at the OPAL reactor (Lucas Heights, NSW, Australia). The USANS measurements were performed to cover a q-range from $3.5 \times 10^{-5}$ Å$^{-1}$ to $10^{-2}$ Å$^{-1}$ using a Bonse-Hart instrument. The samples were loaded into demountable sample cells with a path length of 1 mm. A thermally controlled sample charger was used to control the sample temperature to 25 °C. The experimental USANS data were de-smeared using the Lake algorithm incorporated in NIST USANS macros[72].

**Cryo-scanning electron microscopy:** Samples were placed in a Leica HMP-100 High-Pressure Freezer and stored in liquid nitrogen prior to being loaded into a Leica VCT 100 cryo-transfer shuttle and coated with 7 nm of Pt/C and 5 nm of carbon within a Leica ACE 600 high-vacuum sputter coater. After coating, samples were kept frozen and loaded with the shuttle into a Hitachi S4800-II cold field emission gun scanning electron microscope fitted with a Leica cryo-stage and imaged at 5 kV with an approximate working distance of 10 mm. See **Figure S3**.

**Cryo-transmission electron microscopy:** Samples were vitrified in liquid ethane with a Vitrobot Mark IV (Thermo Scientific). Liquid samples were pipetted onto Quantifoil R2/2 copper grids (Electron Microscopy Sciences) that had been glow-discharged in air (PELCO easiGlow, Ted Pella, Inc.). Gel samples were deposited into the grids by dipping the grid into the gel before blotting and plunge freezing. Grids were transferred into a transmission electron microscope single tilt cryo-holder (Gatan). Grids were images in a Talos L120C (Thermo Scientific) with a high tension of 120 kV with a 4×4 k BM-Ceta CMOS camera. At least 50 images of each sample were taken at magnifications of 28,000X, 57,000X, and 120,000X, yielding a pixel size of 510, 249, and 121 pm, respectively.

**Dilution Studies:** Gelled samples (~ 4.50 g) were transferred to vials to which 7.5 mL DI water was gently added. The water volume was sufficient to bring the highest concentration from a gel to sol state once the gel melted completely. The vials were closed and left undisturbed for 150 days. Photographs were taken at 0, 2, 47, and 150 days.

**References:**

1.	Murray, H. H., Overview — clay mineral applications. *Applied Clay Science* **1991,** *5* (5), 379-395.
2.	Ho, T. A.; Criscenti, L. J.; Greathouse, J. A., Revealing Transition States during the Hydration of Clay Minerals. *The Journal of Physical Chemistry Letters* **2019,** *10* (13), 3704-3709.
3.	Jerolmack, D. J.; Daniels, K. E., Viewing Earth's surface as a soft-matter landscape. *Nature Reviews Physics* **2019,** *1* (12), 716-730.
4.	Ball, P., Shaped from clay. *Nature* **2005**.
5.	Ferris, J. P., Montmorillonite-catalysed formation of RNA oligomers: the possible role of catalysis in the origins of life. *Philosophical Transactions of the Royal Society B: Biological Sciences* **2006,** *361* (1474), 1777-1786.
6.	Langmuir, I., The Role of Attractive and Repulsive Forces in the Formation of Tactoids, Thixotropic Gels, Protein Crystals and Coacervates. *The Journal of Chemical Physics* **1938,** *6* (12), 873-896.
7.	Freundlich, H., Ueber Thixotropie. *Kolloid-Zeitschrift* **1928,** *46* (4), 289-299.
8.	Whittaker, M. L.; Lammers, L. N.; Carrero, S.; Gilbert, B.; Banfield, J. F., Ion exchange selectivity in clay is controlled by nanoscale chemical–mechanical coupling. *Proceedings of the National Academy of Sciences* **2019,** *116* (44), 22052-22057.
9.	Norrish, K., Crystalline Swelling of Montmorillonite: Manner of Swelling of Montmorillonite. *Nature* **1954,** *173* (4397), 256-257.
10.	Swartzen-Allen, S. L.; Matijevic, E., Surface and colloid chemistry of clays. *Chemical Reviews* **1974,** *74* (3), 385-400.
11.	Norrish, K.; Quirk, J. P., Crystalline Swelling of Montmorillonite: Use of Electrolytes to Control Swelling. *Nature* **1954,** *173* (4397), 255-256.
12.	Paineau, E.; Philippe, A. M.; Antonova, K.; Bihannic, I.; Davidson, P.; Dozov, I.; Gabriel, J. C. P.; Impéror-Clerc, M.; Levitz, P.; Meneau, F.; Michot, L. J., Liquid–crystalline properties of aqueous suspensions of natural clay nanosheets. *Liquid Crystals Reviews* **2013,** *1* (2), 110-126.
13.	Michot, L. J.; Paineau, E.; Bihannic, I.; Maddi, S.; Duval, J. F. L.; Baravian, C.; Davidson, P.; Levitz, P., Isotropic/nematic and sol/gel transitions in aqueous suspensions of size selected nontronite NAu1. *Clay Minerals* **2013,** *48* (5), 663-685.
14.	Michot, L. J.; Bihannic, I.; Maddi, S.; Funari, S. S.; Baravian, C.; Levitz, P.; Davidson, P., Liquid–crystalline aqueous clay suspensions. *Proceedings of the National Academy of Sciences* **2006,** *103* (44), 16101-16104.
15.	Michot, L. J.; Bihannic, I.; Maddi, S.; Baravian, C.; Levitz, P.; Davidson, P., Sol/Gel and Isotropic/Nematic Transitions in Aqueous Suspensions of Natural Nontronite Clay. Influence of Particle Anisotropy. 1. Features of the I/N Transition. *Langmuir* **2008,** *24* (7), 3127-3139.
16.	Paineau, E.; Dozov, I.; Philippe, A.-M.; Bihannic, I.; Meneau, F.; Baravian, C.; Michot, L. J.; Davidson, P., In-situ SAXS Study of Aqueous Clay Suspensions Submitted to Alternating Current Electric Fields. *The Journal of Physical Chemistry B* **2012,** *116* (45), 13516-13524.
17.	Bihannic, I.; Michot, L. J.; Lartiges, B. S.; Vantelon, D.; Labille, J.; Thomas, F.; Susini, J.; Salomé, M.; Fayard, B., First Direct Visualization of Oriented Mesostructures in Clay Gels by Synchrotron-Based X-ray Fluorescence Microscopy. *Langmuir* **2001,** *17* (14), 4144-4147.
18.	Davidson, P.; Gabriel, J.-C. P., Mineral liquid crystals. *Current Opinion in Colloid & Interface Science* **2005,** *9* (6), 377-383.
19.	Mourad, M. C. D.; Wijnhoven, J. E. G. J.; van 't Zand, D. D.; van der Beek, D.; Lekkerkerker, H. N. W., Gelation versus liquid crystal phase transitions in suspensions of plate-like particles. *Philosophical Transactions of the Royal Society A: Mathematical, Physical and Engineering Sciences* **2006,** *364* (1847), 2807-2816.

20. Van Olphen, H., Rheological phenomena of clay sols in connection with the charge distribution on the micelles. *Discussions of the Faraday Society* **1951,** *11* (0), 82-84.
21. Morvan, M.; Espinat, D.; Lambard, J.; Zemb, T., Ultrasmall- and small-angle X-ray scattering of smectite clay suspensions. *Colloids and Surfaces A: Physicochemical and Engineering Aspects* **1994,** *82* (2), 193-203.
22. Norrish, K., The swelling of montmorillonite. *Discussions of the Faraday Society* **1954,** *18* (0), 120-134.
23. Callaghan, I. C.; Ottewill, R. H., Interparticle forces in montmorillonite gels. *Faraday Discussions of the Chemical Society* **1974,** *57* (0), 110-118.
24. Paineau, E.; Bihannic, I.; Baravian, C.; Philippe, A.-M.; Davidson, P.; Levitz, P.; Funari, S. S.; Rochas, C.; Michot, L. J., Aqueous Suspensions of Natural Swelling Clay Minerals. 1. Structure and Electrostatic Interactions. *Langmuir* **2011,** *27* (9), 5562-5573.
25. Abend, S.; Lagaly, G., Sol–gel transitions of sodium montmorillonite dispersions. *Applied Clay Science* **2000,** *16* (3), 201-227.
26. Sposito, G.; Prost, R., Structure of water adsorbed on smectites. *Chemical Reviews* **1982,** *82* (6), 553-573.
27. Shalkevich, A.; Stradner, A.; Bhat, S. K.; Muller, F.; Schurtenberger, P., Cluster, Glass, and Gel Formation and Viscoelastic Phase Separation in Aqueous Clay Suspensions. *Langmuir* **2007,** *23* (7), 3570-3580.
28. Michot, L. J.; Bihannic, I.; Porsch, K.; Maddi, S.; Baravian, C.; Mougel, J.; Levitz, P., Phase Diagrams of Wyoming Na-Montmorillonite Clay. Influence of Particle Anisotropy. *Langmuir* **2004,** *20* (25), 10829-10837.
29. Mourchid, A.; Lécolier, E.; Van Damme, H.; Levitz, P., On Viscoelastic, Birefringent, and Swelling Properties of Laponite Clay Suspensions: Revisited Phase Diagram. *Langmuir* **1998,** *14* (17), 4718-4723.
30. Michot, L. J.; Baravian, C.; Bihannic, I.; Maddi, S.; Moyne, C.; Duval, J. F. L.; Levitz, P.; Davidson, P., Sol−Gel and Isotropic/Nematic Transitions in Aqueous Suspensions of Natural Nontronite Clay. Influence of Particle Anisotropy. 2. Gel Structure and Mechanical Properties. *Langmuir* **2009,** *25* (1), 127-139.
31. Goyal, A.; Palaia, I.; Ioannidou, K.; Ulm, F.-J.; van Damme, H.; Pellenq, R. J.-M.; Trizac, E.; Del Gado, E., The physics of cement cohesion. *Science Advances* **2021,** *7* (32), eabg5882.
32. Plassard, C.; Lesniewska, E.; Pochard, I.; Nonat, A., Nanoscale Experimental Investigation of Particle Interactions at the Origin of the Cohesion of Cement. *Langmuir* **2005,** *21* (16), 7263-7270.
33. Bailey, L.; Lekkerkerker, H. N. W.; Maitland, G. C., Smectite clay – inorganic nanoparticle mixed suspensions: phase behaviour and rheology. *Soft Matter* **2015,** *11* (2), 222-236.
34. McBride, M. B.; Baveye, P., Diffuse Double-Layer Models, Long-Range Forces, and Ordering in Clay Colloids. *Soil Science Society of America Journal* **2002,** *66* (4), 1207-1217.
35. Quirk, J. P., Comments on "Diffuse double-layer models, long-range forces, and ordering of clay colloids". *Soil Science Society of America Journal* **2003,** *67* (6), 1960-1961.
36. Quirk, J. P., Interparticle Forces: A Basis for the Interpretation of Soil Physical Behavior. In *Advances in Agronomy*, Sparks, D. L., Ed. Academic Press: 1994; Vol. 53, pp 121-183.
37. Overbeek, J. T. G., On the interaction of highly charged plates in an electrolyte: a correction. *Molecular Physics* **1993,** *80* (3), 685-694.
38. Smalley, M. V.; Sogami, I. S., On the interaction of highly charged plates in an electrolyte. *Molecular Physics* **1995,** *85* (5), 869-881.
39. Batista, C. A. S.; Larson, R. G.; Kotov, N. A., Nonadditivity of nanoparticle interactions. *Science* **2015,** *350* (6257), 1242477.
40. Ilavsky, J.; Jemian, P. R.; Allen, A. J.; Zhang, F.; Levine, L. E.; Long, G. G., Ultra-small-angle X-ray scattering at the Advanced Photon Source. *Journal of Applied Crystallography* **2009,** *42* (3), 469-479.

41. Jönsson, B.; Persello, J.; Li, J.; Cabane, B., Equation of State of Colloidal Dispersions. *Langmuir* **2011,** *27* (11), 6606-6614.
42. Bonnet-Gonnet, C.; Belloni, L.; Cabane, B., Osmotic Pressure of Latex Dispersions. *Langmuir* **1994,** *10* (11), 4012-4021.
43. Calabrese, V.; Haward, S. J.; Shen, A. Q., Effects of Shearing and Extensional Flows on the Alignment of Colloidal Rods. *Macromolecules* **2021,** *54* (9), 4176-4185.
44. Michot, L. J.; Bihannic, I.; Thomas, F.; Lartiges, B. S.; Waldvogel, Y.; Caillet, C.; Thieme, J.; Funari, S. S.; Levitz, P., Coagulation of Na-Montmorillonite by Inorganic Cations at Neutral pH. A Combined Transmission X-ray Microscopy, Small Angle and Wide Angle X-ray Scattering Study. *Langmuir* **2013,** *29* (10), 3500-3510.
45. Murphy, R. P.; Hong, K.; Wagner, N. J., Thermoreversible Gels Composed of Colloidal Silica Rods with Short-Range Attractions. *Langmuir* **2016,** *32* (33), 8424-8435.
46. Kishore, S.; Srivastava, S.; Bhatia, S. R., Microstructure of colloid-polymer mixtures containing charged colloidal disks and weakly-adsorbing polymers. *Polymer* **2016,** *105*, 461-471.
47. Bhatia, S. R., Ultra-small-angle scattering studies of complex fluids. *Current Opinion in Colloid & Interface Science* **2005,** *9* (6), 404-411.
48. Dimon, P.; Sinha, S. K.; Weitz, D. A.; Safinya, C. R.; Smith, G. S.; Varady, W. A.; Lindsay, H. M., Structure of Aggregated Gold Colloids. *Physical Review Letters* **1986,** *57* (5), 595-598.
49. Shan, L.; Xie, R.; Wagner, N. J.; He, H.; Liu, Y., Microstructure of neat and SBS modified asphalt binder by small-angle neutron scattering. *Fuel* **2019,** *253*, 1589-1596.
50. Gabriel, J.-C. P.; Camerel, F.; Lemaire, B. J.; Desvaux, H.; Davidson, P.; Batail, P., Swollen liquid-crystalline lamellar phase based on extended solid-like sheets. *Nature* **2001,** *413* (6855), 504-508.
51. Segad, M.; Cabane, B.; Jönsson, B., Reactivity, swelling and aggregation of mixed-size silicate nanoplatelets. *Nanoscale* **2015,** *7* (39), 16290-16297.
52. Mouzon, J.; Bhuiyan, I. U.; Hedlund, J., The structure of montmorillonite gels revealed by sequential cryo-XHR-SEM imaging. *Journal of Colloid and Interface Science* **2016,** *465*, 58-66.
53. Deirieh, A.; Chang, I. Y.; Whittaker, M. L.; Weigand, S.; Keane, D.; Rix, J.; Germaine, J. T.; Joester, D.; Flemings, P. B., Particle arrangements in clay slurries: The case against the honeycomb structure. *Applied Clay Science* **2018,** *152*, 166-172.
54. Hotton, C.; Sirieix-Plénet, J.; Ducouret, G.; Bizien, T.; Chennevière, A.; Porcar, L.; Michot, L.; Malikova, N., Organisation of clay nanoplatelets in a polyelectrolyte-based hydrogel. *Journal of Colloid and Interface Science* **2021,** *604*, 358-367.
55. Ruzicka, B.; Zulian, L.; Zaccarelli, E.; Angelini, R.; Sztucki, M.; Moussaïd, A.; Ruocco, G., Competing Interactions in Arrested States of Colloidal Clays. *Physical Review Letters* **2010,** *104* (8), 085701.
56. Angelini, R.; Zaccarelli, E.; de Melo Marques, F. A.; Sztucki, M.; Fluerasu, A.; Ruocco, G.; Ruzicka, B., Glass–glass transition during aging of a colloidal clay. *Nature Communications* **2014,** *5* (1), 4049.
57. Ramsay, J. D.; Swanton, S. W.; Bunce, J., Swelling and dispersion of smectite clay colloids: determination of structure by neutron diffraction and small-angle neutron scattering. *Journal of the Chemical Society, Faraday Transactions* **1990,** *86* (23), 3919-3926.
58. Segad, M.; Hanski, S.; Olsson, U.; Ruokolainen, J.; Åkesson, T.; Jönsson, B., Microstructural and Swelling Properties of Ca and Na Montmorillonite: (In Situ) Observations with Cryo-TEM and SAXS. *The Journal of Physical Chemistry C* **2012,** *116* (13), 7596-7601.
59. Segad, M.; Jönsson, B.; Cabane, B., Tactoid Formation in Montmorillonite. *The Journal of Physical Chemistry C* **2012,** *116* (48), 25425-25433.
60. Ramsay, J. D. F.; Lindner, P., Small-angle neutron scattering investigations of the structure of thixotropic dispersions of smectite clay colloids. *Journal of the Chemical Society, Faraday Transactions* **1993,** *89* (23), 4207-4214.

61.     Lekkerkerker, H. N. W.; Vroege, G. J., Liquid crystal phase transitions in suspensions of mineral colloids: new life from old roots. *Philosophical Transactions of the Royal Society A: Mathematical, Physical and Engineering Sciences* **2013,** *371* (1988), 20120263.
62.     Silbert, G.; Ben-Yaakov, D.; Dror, Y.; Perkin, S.; Kampf, N.; Klein, J., Long-Ranged Attraction between Disordered Heterogeneous Surfaces. *Physical Review Letters* **2012,** *109* (16), 168305.
63.     Adar, R. M.; Andelman, D.; Diamant, H., Electrostatics of patchy surfaces. *Advances in Colloid and Interface Science* **2017,** *247*, 198-207.
64.     Shen, X.; Bourg, I. C., Molecular dynamics simulations of the colloidal interaction between smectite clay nanoparticles in liquid water. *Journal of Colloid and Interface Science* **2021,** *584*, 610-621.
65.     French, R. H.; Parsegian, V. A.; Podgornik, R.; Rajter, R. F.; Jagota, A.; Luo, J.; Asthagiri, D.; Chaudhury, M. K.; Chiang, Y.-m.; Granick, S.; Kalinin, S.; Kardar, M.; Kjellander, R.; Langreth, D. C.; Lewis, J.; Lustig, S.; Wesolowski, D.; Wettlaufer, J. S.; Ching, W.-Y.; Finnis, M.; Houlihan, F.; von Lilienfeld, O. A.; van Oss, C. J.; Zemb, T., Long range interactions in nanoscale science. *Reviews of Modern Physics* **2010,** *82* (2), 1887-1944.
66.     Jellander, R.; Marčelja, S.; Quirk, J. P., Attractive double-layer interactions between calcium clay particles. *Journal of Colloid and Interface Science* **1988,** *126* (1), 194-211.
67.     Ilavsky, J.; Zhang, F.; Andrews, R. N.; Kuzmenko, I.; Jemian, P. R.; Levine, L. E.; Allen, A. J., Development of combined microstructure and structure characterization facility for in situ and operando studies at the Advanced Photon Source. *Journal of Applied Crystallography* **2018,** *51* (3), 867-882.
68.     Li, T.; Senesi, A. J.; Lee, B., Small Angle X-ray Scattering for Nanoparticle Research. *Chemical Reviews* **2016,** *116* (18), 11128-11180.
69.     Ilavsky, J., Nika: software for two-dimensional data reduction. *Journal of Applied Crystallography* **2012,** *45* (2), 324-328.
70.     Ilavsky, J.; Jemian, P. R., Irena: tool suite for modeling and analysis of small-angle scattering. *Journal of Applied Crystallography* **2009,** *42* (2), 347-353.
71.     Rehm, C.; Brule, A.; Freund, A. K.; Kennedy, S. J., Kookaburra: the ultra-small-angle neutron scattering instrument at OPALThis article will form part of a virtual special issue of the journal, presenting some highlights of the 15th International Small-Angle Scattering Conference (SAS2012). This special issue will be available in late 2013/early 2014. *Journal of Applied Crystallography* **2013,** *46* (6), 1699-1704.
72.     Kline, S., Reduction and analysis of SANS and USANS data using IGOR Pro. *Journal of Applied Crystallography* **2006,** *39* (6), 895-900.

## Supplementary

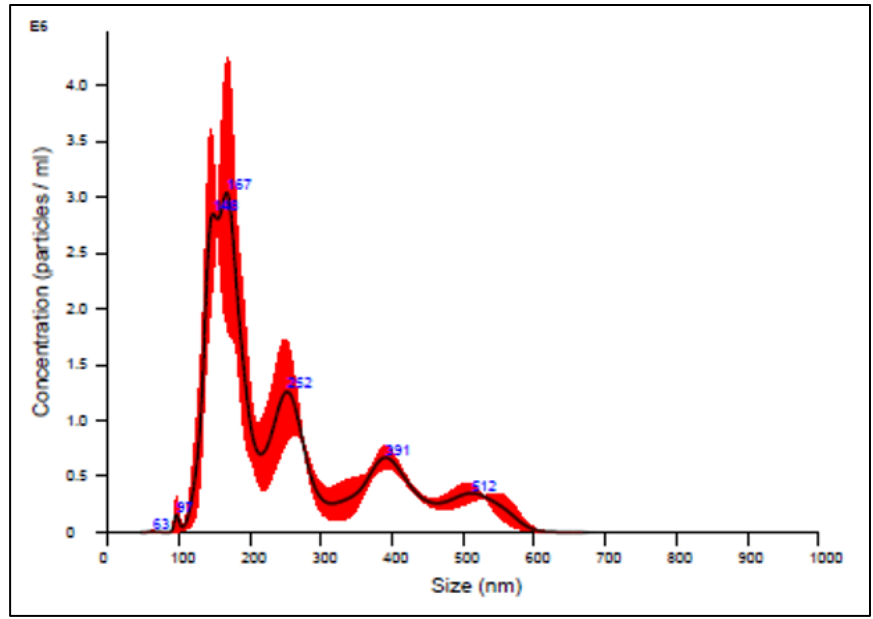

**Figure S1:** Representative dynamic light scattering particle size of sodium-montmorillonite sample. The maximum particle size is in the range of 600 nm.

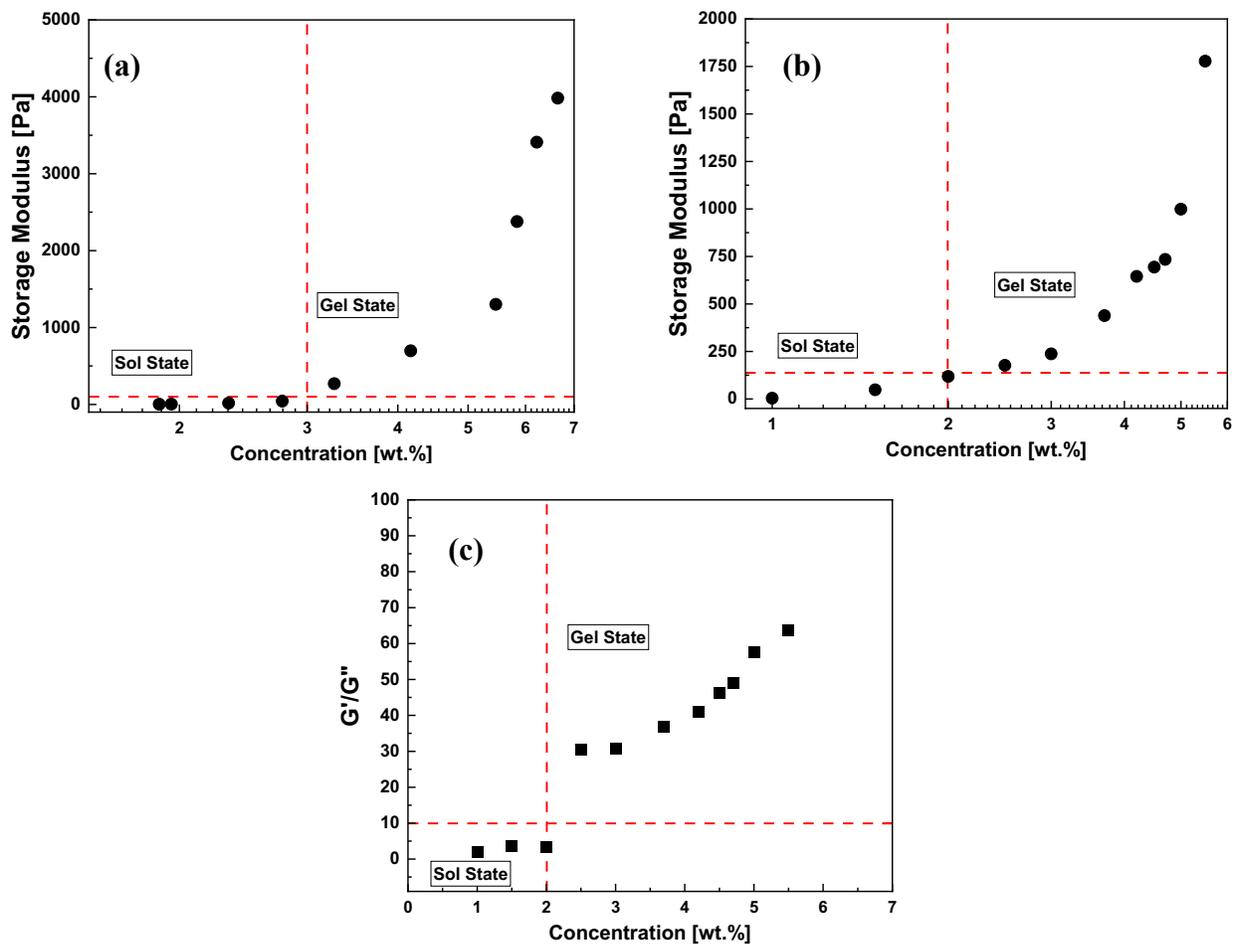

**Figure S2:** (a) Evolution of storage modulus (G') as a function of sodium-montmorillonite concentration. G' increases exponentially as a function of sodium-montmorillonite (Na-Mt) concentration in the gel state (b) G' as a function of Na-Mt concentration in deuterium oxide (D$_2$O), the gel point is between Na-Mt concentration 2.0-2.5 wt.%. (c) Ratio of storage(G') and loss modulus (G'') for Na-Mt suspensions in D$_2$O.

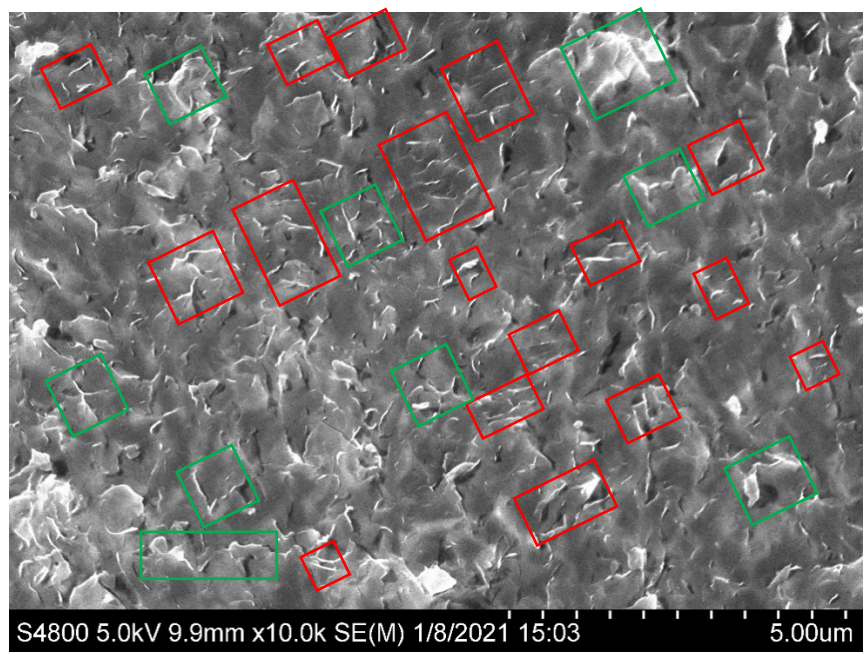

**Figure S3:** Cryo-scanning electron micrograph of montmorillonite in the gel state. The gel sample shows domains of aggregated and ordered entities. Red boxes represent particle-particle ordering, and green boxes represent particle-particle aggregation.

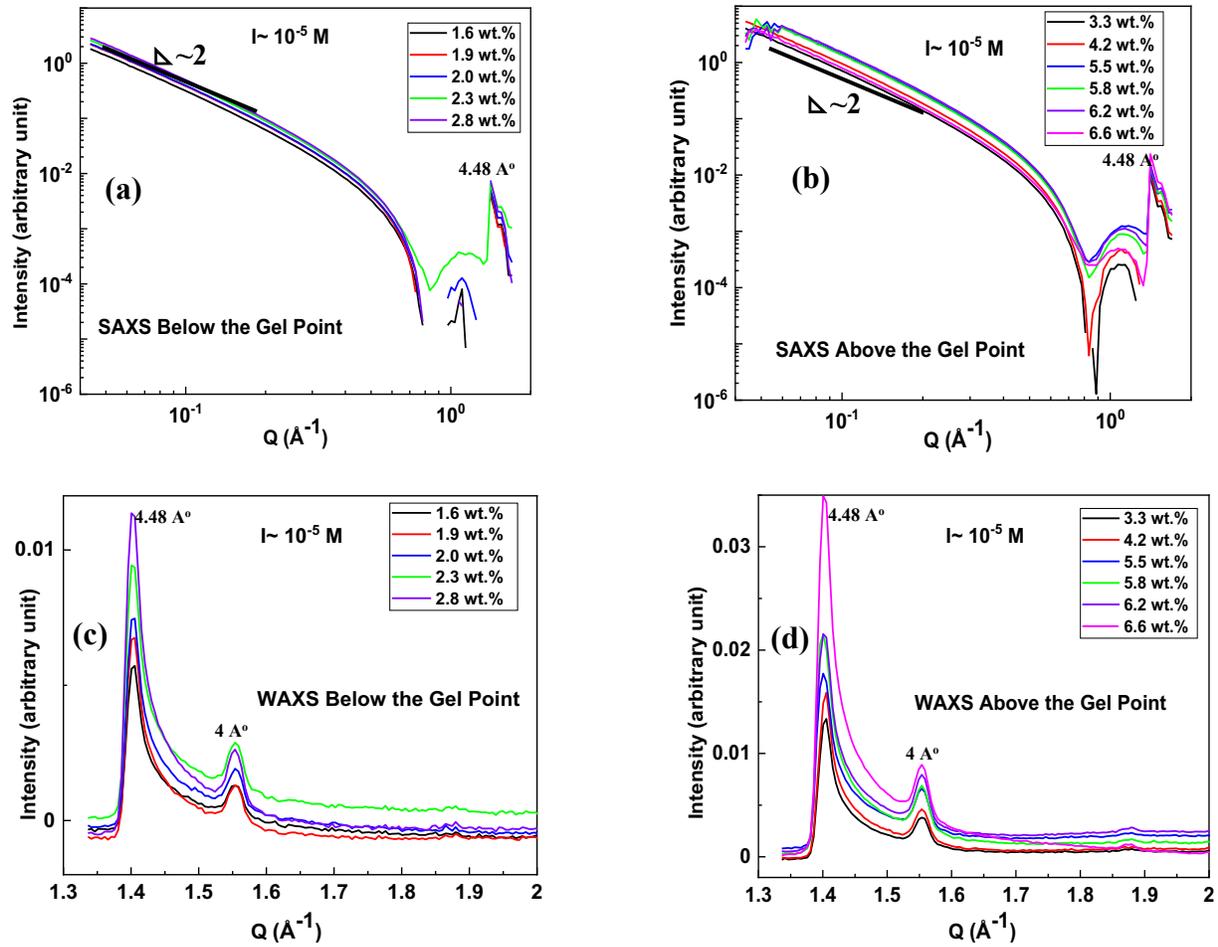

**Figure S4:** (a,b) Small angle X-ray scattering and (c,d) wide angle X-ray scattering of Na-Mt suspensions in (a,c) sol and (b,d) gel states.

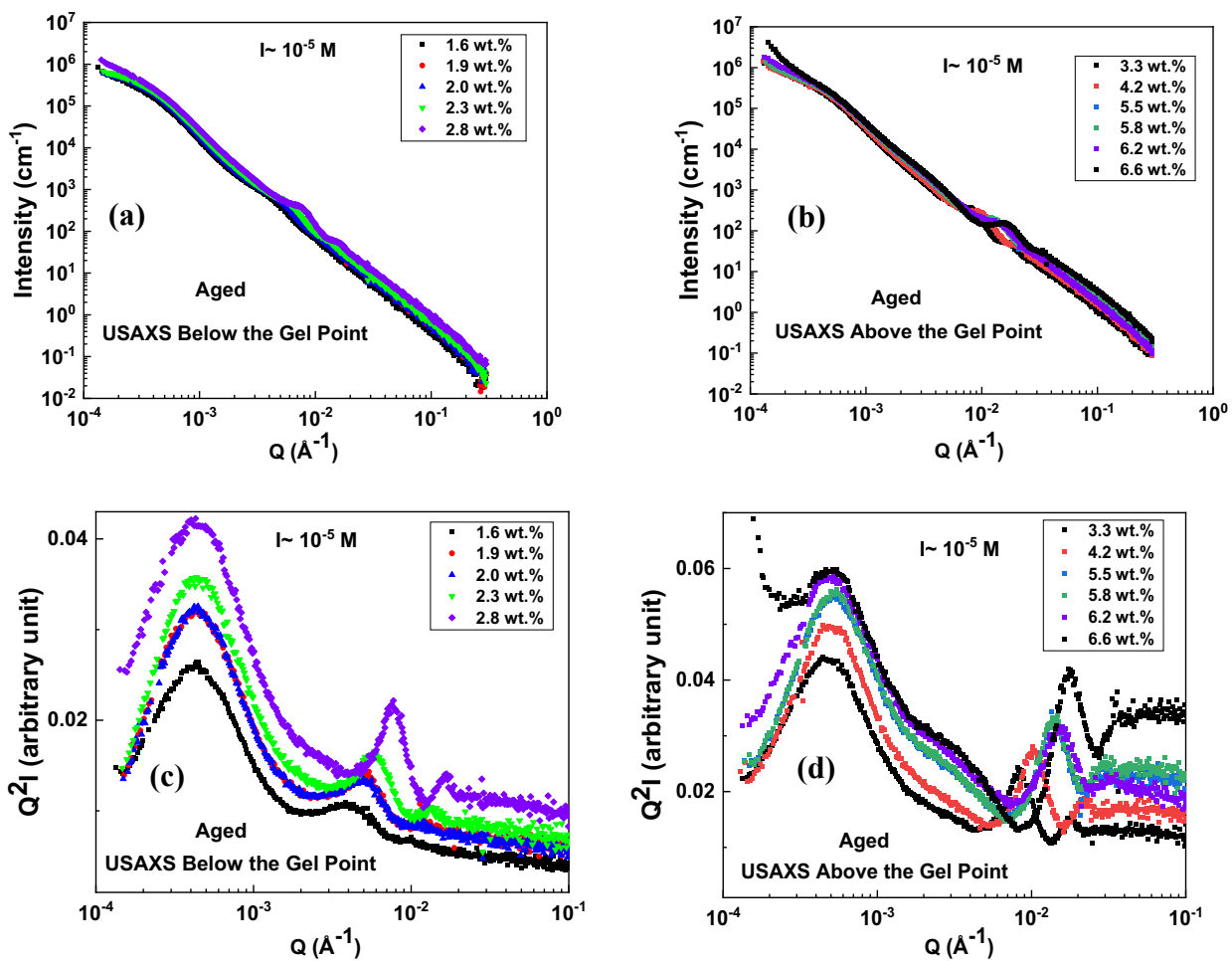

**Figure S5:** Ultra-small angle X-ray scattering of Na-Mt suspensions aged for 3 months: (a, c) in the sol state and (b, d) in the gel state

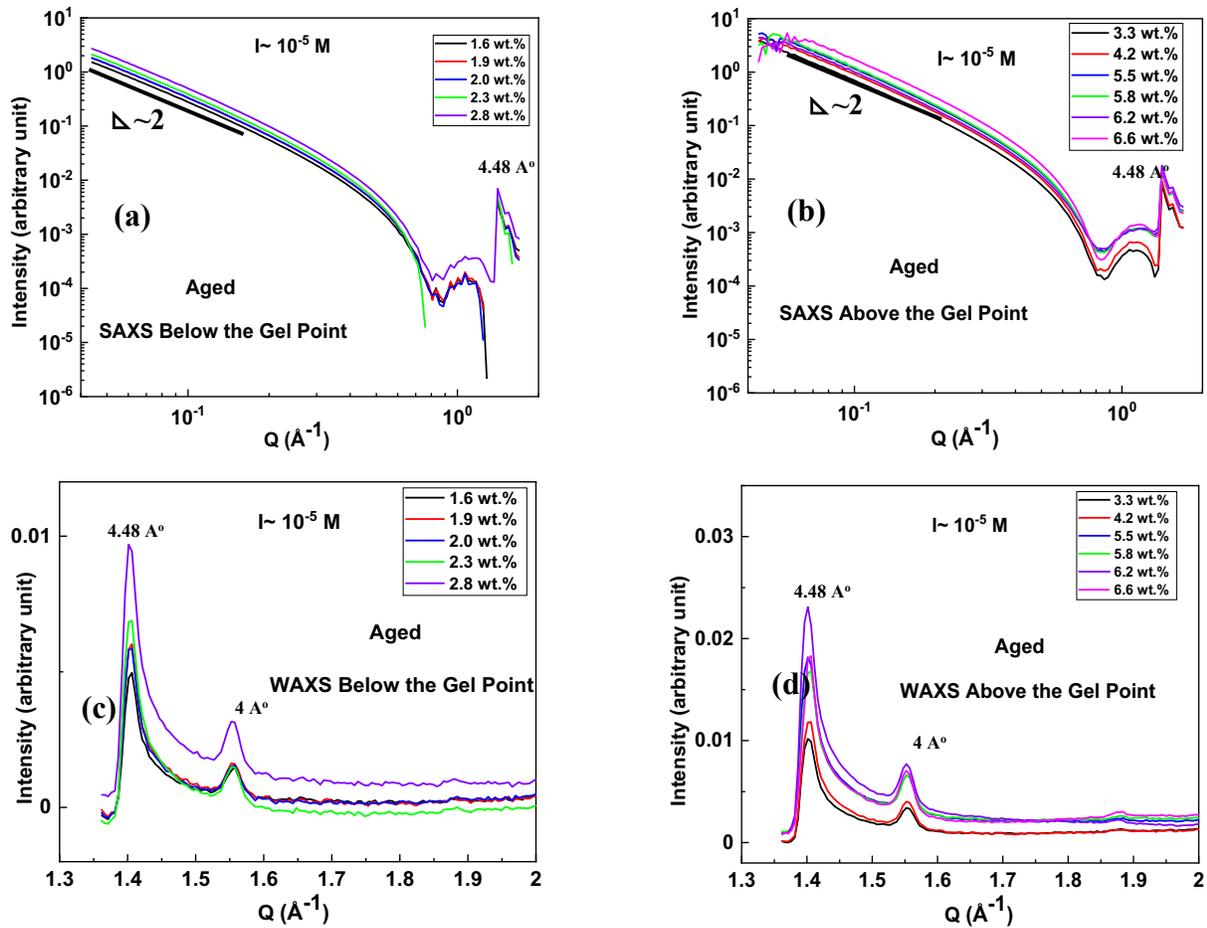

**Figure S6:** (a, b) Small angle X-ray scattering and (c, d) wide angle X-ray scattering of Na-Mt suspensions aged for 3 months: (a, c) in the sol state and (b, d) in the gel state.